\documentstyle[aps]{revtex}
\begin{document}
\draft
\title{Avalanches at rough surfaces}
\author{G C Barker}
\address{Institute of Food Research,\\
Norwich Research Park,\\ Colney, Norwich NR4 7UA, UK\\ email:
barker@bbsrc.ac.uk}
\author{Anita Mehta}
\address{S N Bose National Centre for Basic Sciences\\
Block JD Sector III Salt Lake\\ Calcutta 700091, India\\ email:
anita@boson.bose.res.in}
\maketitle

\begin{abstract}
We describe the surface properties of a simple lattice model of a
sandpile that includes evolving structural disorder. We present a
dynamical scaling hypothesis for generic sandpile automata, and
additionally explore the kinetic roughening of the sandpile
surface,  indicating its relationship with the sandpile evolution.
Finally, we comment on the surprisingly good agreement found
between this model, and a previous continuum model of sandpile
dynamics, from the viewpoint of critical phenomena.
\end{abstract}


Avalanches are the signatures of instabilities on an evolving
surface: regions on a sandpile, for example, which protrude
excessively from the surface, get dislodged by such mechanisms.
This most intuitive picture of avalanching \cite{held,jaegerRMP}
is the one we seek to model and study: although presented here in
the context of sandpiles, a similar picture may be relevant to
intermittent granular flows along an inclined plane \cite{daerr},
or to sediment consolidation \cite{ball}. As deposition occurs on
a sandpile surface, clusters of grains grow unevenly at different
positions and roughness builds up until further deposition renders
some of the clusters unstable. These then start 'toppling', so
that grains from an already unstable cluster flow down the
sandpile, knocking off grains from other similar clusters which
they encounter. The net effect of this is to 'wipe off'
protrusions (where there is a surfeit of grains at a cluster) and
to 'fill in' dips, where the oncoming avalanche can disburse some
of its grains. In short, the surface is smoothed by the passage of
the avalanche so that there is a rough precursor surface, and a
smoothed post-avalanche surface.

In earlier work \cite{amcoupled1}, \cite{amcoupled2}, we have
explored this issue via an analytical model involving coupled
continuum equations. Here we use a cellular-automaton model
\cite{amca} of an evolving sandpile to look in more depth at the
mechanisms by which  a large avalanche smooths the surface. Our
sandpile model is a 'disordered' and nonabelian version of the
basic Kadanoff cellular automaton \cite{kadanoff}; a further
degree of freedom, that involves granular reorganisation within
columns, is added to the basic model which includes only granular
flow between columns. In this sense, each column is regarded as a
cluster of grains so that we represent  intracluster as well as
intercluster grain relaxations, in accord with a previous
understanding of sandpile dynamics \cite{ambook}.

Our disordered model sandpile \cite{caparam} is built from
rectangular lattice grains, that have aspect ratio $a \le 1$
arranged in columns $i$ with $ 1 \le i \le L$, where $L$ is the
system size. Each grain is labelled by its column index $i$  and
by an orientational index $0$ or $1$, corresponding respectively
to whether the grain rests on its larger or smaller edge.

The dynamics of our model have been described at length elsewhere
\cite{amca} but we review the essentials here:
\begin{itemize}
\item
Grains are deposited on the sandpile with fixed probabilities of
landing in the $0$ or $1$ position.
\item The incoming grains, as well as all the grains in the same column,
can then 'flip' to the other orientation stochastically (with
probabilities which decrease with depth from the surface). This
'flip', or change of orientation, is our simple representation of
{\it collective dynamics in granular clusters} since typically
clusters reorganise owing to the slight orientational movements of
the grains within them \cite{ambook}.
\item
Column heights are then computed as follows: the height of column
$i$ at time $t$, $h(i,t)$, can be expressed in terms of the
instantaneous numbers of $0$ and $1$ grains, $ n_0(i,t)$ and
$n_1(i,t)$ respectively:
\begin{equation}
        h(i,t) = n_1(i,t) + a n_0(i,t)
\end{equation}
\item
Finally, grains fall to the next column down the sandpile
(maintaining their orientation as they do so) if the height
difference exceeds a specified threshold in the normal way
\cite{kadanoff} (the pile is local, limited and has a fall number
of two). At this point, avalanching occurs.
\end{itemize}

In earlier papers, \cite{amca}, it was shown that the presence of
disorder led, for large enough system sizes, to a preferred size
of large avalanches. In the absence of disorder, scale-invariant
avalanche statistics were observed. In this work, we focus on the
state of the sandpile surface \cite{ampune} in a bid to correlate
its evolution with the onset and propagation of avalanches.

\section {Dynamical scaling for sandpile cellular automata}
It is customary in the study of generalised surfaces to examine
the widths generated by kinetic roughening \cite{krug}, and then
establish properties related to {\it dynamical scaling}. However,
the kinetic roughening of sandpile cellular automata has never
been investigated; to begin with, therefore we postulate a
principle of {\it dynamical scaling for sandpile cellular
automata} in terms of the surface width $W$ of the sandpile
automaton:

\begin{eqnarray}
 W(t) &\sim&  t^{\beta},  t \ll t_{crossover} \equiv L^z
\label{betaeqn}\\
 W(L) &\sim&  L^{\alpha}, L \to \infty \label{alphaeqn}
\end{eqnarray}

As in the case of interfacial widths, these equations signify the
following sequence of roughening regimes:
\begin{enumerate}
\item To start with, roughening occurs at the CA sandpile surface
in a time-dependent way; after an initial transient, the width
scales asymptotically with time $t$ as $t^{\beta}$, where $\beta$
is the {\it temporal roughening} exponent. This regime is
appropriate for all times less than the crossover time
$t_{crossover} \equiv L^z$, where $z$ = $\alpha/\beta$ is the
dynamical exponent and $L$ the system size.
\item After the surface has {\it saturated}, i.e. its width no longer
grows with time, the {\it spatial roughening} characteristics of
the mature interface can be measured in terms of $\alpha$, an
exponent characterising the dependence of the width on  $L$.
\end{enumerate}

We define the  surface width $W(t)$ for a sandpile automaton in
terms of the mean-squared deviations from a suitably defined mean
surface; in  analogy with the  conventional counterpart for
interface growth \cite{krug}, we define the instantaneous mean
surface of a sandpile automaton as the surface about which the sum
of column {\it height} fluctuations vanishes. Clearly, in an
evolving surface, this must be a function of time; hence all
quantities in the following analysis  will be presumed to be
instantaneous.

The mean slope $<s(t)>$ defines  expected column heights,
$h_{av}(i,t)$, according to
\begin{equation}
            h_{av}(i,t) = i <s(t)>
\end{equation}
where we have assumed that column $1$ is at the bottom of the
pile. Column height deviations are defined by
\begin{equation}
        dh(i,t) = h(i,t) - h_{av}(i,t)  =  h(i,t) - i <s(t)>
\end{equation}
The mean slope must therefore satisfy
\begin{equation}
            \Sigma_i[ h(i,t) - i <s(t)> ] = 0
\end{equation}
since the instantaneous deviations about it vanish; thus
\begin{equation}
            <s(t)> = 2 \Sigma_i[h(i,t)] / L(L+1)
\end{equation}
(We note that this slope is distinct from the quantity $<s'(t)> =
h(L,t)/L$ that is obtained from the average of all the local
slopes $s(i,t) = h(i,t) - h(i-1,t)$, about which {\it slope}
fluctuations would vanish on average).

The instantaneous width of the surface of a sandpile automaton,
$W(t)$, can  be defined as:
\begin{equation}
            W(t) = \sqrt{ \Sigma_i[dh(i,t)^2] / L }
\end{equation}
which can in turn be averaged over several realizations to give,
$<W>$, the average surface width in the steady state.

We also compute here the height-height correlation function,
$ C(j,t)$, which is defined by
\begin{equation}
            C(j,t) = <dh(i,t)dh(i+j,t)> / <dh(i,t)^2>
            \label{hheqn}
\end{equation}
where the mean values are evaluated over all pairs of surface
sites separated by $j$ lattice spacings:
\begin{equation}
    <dh(i,t)dh(i+j,t)> = \Sigma_i(dh(i,t)dh(i+j,t)) / (L-j)
\end{equation}
for $0 \le j < L$. This function is symmetric and  can be averaged
over several realizations to give the average correlation function
$<C(j)>$.

\section{Qualitative effects of avalanching on surfaces}
Before moving on to the quantitative descriptors of sandpile
avalanching and surface roughening, we present some results using
more qualitative indicators. Recent experiments \cite{daerr} on
sandpile avalanches have indicated that there are at least two
broad categories;  'uphill' avalanches, which are typically large,
and  'triangular' avalanches which are generally smaller in size.
We have found evidence of this in a $(2+1)d$ disordered model of
sandpile avalanches, which will be presented elsewhere
\cite{amunpub}; but in this work we discuss analogues in $(1+1)d$,
which are respectively 'wedge-shaped' and 'flat' avalanches. The
following data indicate that it is the larger wedge-shaped
avalanches which alter surface slope and width, while the flatter,
smaller avalanches alter neither very much. This is in accord with
earlier work, where it was found that larger avalanches are the
consequence of accumulated disorder, while the smaller ones can
cause disordered regions to build up along the sandpile surface
\cite{ampre}.

Figure \ref{Fig1}(a)  shows a time series for the mass of a large
($L=256$) evolving disordered sandpile automaton. The series has a
typical quasiperiodicity \cite{held}. The vertical line denotes
the position of a particular 'large' event, while Figure
\ref{Fig1}(b) shows the avalanche size distribution for the
sandpile. Note the peak, corresponding to the preferred large
avalanches, which was analysed extensively in earlier work
\cite{amca}. Our data shows that the avalanche highlighted in
Figure \ref{Fig1}(a) drained off approximately $5$ per cent of the
mass of the sandpile,  placing it close to the 'second peak' of
Figure \ref{Fig1}(b). Figure \ref{Fig1}(c) shows the outline of
the full avalanche before and after this event with its initiation
site marked by an arrow; we note that, as is often the case in one
dimension, the avalanche is 'uphill'. The inset shows the relative
motion of the surface during this event; we note that the
signatures of smoothing by avalanches are already evident as the
precursor state in the inset is much rougher than the final state.
Finally we show in Figure \ref{Fig1}(d) the grain-by-grain picture
of the aftermath pile superposed on the precursor pile, which is
shown in shadow. An examination of the aftermath pile and the
precursor pile shows that the propagation of the avalanche across
the  upper half of the pile has left only a very few disordered
sites in its wake (i.e. the majority of the remaining sites are
$0$ type) whereas the lower half (which was undisturbed by the
avalanche) still contains many disordered, i.e. $1$ type sites in
the boundary layer. This leads us to suggest that the larger
avalanches rid the boundary layer of its disorder-induced
roughness, a fact that is borne out by our more quantitative
investigations.

In fact, our studies have revealed that the very largest
avalanches, which are system-spanning, remove virtually all
disordered sites from the surface layer; one is then left with a
normal 'ordered' sandpile, where the avalanches have their usual
scaling form for as long as it takes for a layer of disorder to
build up. When the disordered layer reaches a critical size,
another large event is unleashed; this is the underlying reason
for the quasiperiodic form of the time series shown in Figure
\ref{Fig1}(a).

Before moving on to more quantitative features, we show for
comparison the sequence of Figure \ref{Fig1}, for
\begin{enumerate}
\item an ordered pile - Figure \ref{Fig2}(a-d)
\item a small disordered pile - Figure \ref{Fig3}(a-d)
\end{enumerate}

We note the following features:
\begin{itemize}
\item The small disordered pile has a mass time series (Figure
\ref{Fig3}(a)) that
is midway between the scale-invariance of the ordered pile (Figure
\ref{Fig2}(a)) and the quasiperiodicity of the large disordered
pile (Figure \ref{Fig1}(a)).
\item The avalanche size distribution of the small disordered pile
(Figure \ref{Fig3}(b)) is likewise intermediate between that of
the ordered pile (which shows the scale invariance observed by
Kadanoff et al. \cite{kadanoff}) and the two-peaked distribution
characteristic of the disordered pile \cite{amca}.
\item In both small and large disordered piles, we see evidence
of large 'uphill' avalanches which shave off  a thick boundary
layer containing large numbers of disordered sites, and leave
behind a largely ordered pile (see Figure \ref{Fig1}(c-d) and
Figure \ref{Fig3}(c-d)). By contrast the ordered pile loses
typically two commensurate layers even in the largest avalanche,
with a correspondingly unexciting aftermath state left behind in
its wake (Figure \ref{Fig2}(c-d)).
\end{itemize}

We conclude from this that there is, even at a qualitative level,
a post-avalanche smoothing of the sandpile surface, beyond a
crossover length, as found in earlier  work on continuum models
\cite{amcoupled1}; importantly, our discrete model  reveals that
this is achieved by the removal of (orientational) disorder, the
implications of which we will discuss in our concluding section.
The existence of the  crossover length, in terms of the mass time
series, has also been observed in experiment \cite{held}.

\section{Quantitative effects of avalanching on surfaces}

\subsection{Intrinsic properties of sandpile surfaces}
Inspired by the picture of smoothing avalanches, we have
investigated many of the material properties of the sandpile in
the special pre- and post-avalanche configurations. From these we
have drawn the following conclusions:
\begin{itemize}

\item
The {\it mean slope} of the disordered sandpile peaks (see Table
\ref{Table1}) before a large avalanche and drops immediately
after; this statement is true for events of any size and thus
remains trivially true for the ordered sandpile.
\item
The {\it packing fraction} $\phi$ of the disordered sandpile increases
after a large event, i.e. effective consolidation occurs during
avalanching (see Table \ref{Table1}). This consolidation via
avalanching mirrors that which occurs when a sandpile is shaken
with low-intensity vibrations \cite{amprl,nagel}.
\item
However, a far deeper statement can be made about the comparison
of the surface width for  pre- and post- large event sandpiles;
Table \ref{Table1} shows that the surface width goes down
considerably during an event, once again suggesting that a rough
precursor pile is smoothed by the propagation of a large
avalanche.
\end{itemize}

We have  also investigated the dependence of various material
properties of a disordered sandpile on the aspect ratio of the
grains \cite{frette}. Table \ref{Table2} shows our results, and
Figure \ref{Fig4} illustrates the variation of the avalanche size
distribution.

There is a transition as aspect ratios of $0.7$ are approached
from above or below; we have shown above that piles with these
'critical' aspect ratios manifest strong disorder \cite{amca} in
the sense of:
\begin{itemize}
\item a 'second peak' in the avalanche size distribution
denoting a preferred size of large avalanches
\item large surface widths denoting an increased surface
roughness
\item a strong correlation between interfacial roughness
and avalanche flow since the mean surface width varies
dramatically in the pre- and post- large event piles.
\end{itemize}

Clearly, sandpiles containing grains with aspect ratios close to
unity act essentially as totally ordered piles \cite{kadanoff};
there is however a significant symmetry in the shape of the
avalanche size distribution curves above and below the transition
region (see Figure \ref{Fig4}(a) and (d)). These size
distributions are reminiscent of those obtained in earlier work
\cite{amca} for the case of 'uniform disorder' (which referred to
piles that have disorder throughout their volume  rather than, as
is the present case, disorder concentrated in a boundary layer).
These observations lead us to speculate that there exist at least
three types of avalanche spectra:
\begin{enumerate}
\item the scale-invariant statistics characteristic of ordered
sandpiles
\item the strongly disordered statistics characterized
by a second peak in the distribution; which we have
obtained for specific values of the aspect ratio
in the case where the disorder is concentrated in a boundary layer
\item the more weakly disordered region (characterized by
a flatter size distribution of avalanche sizes which is, nevertheless,
{\it not} scale-invariant) obtained in the intermediate regimes
of aspect ratio (as well as in the case of uniform disorder).
\end{enumerate}

It is clear  that {\it the presence of inherent inhomogeneities in
grain shape (which we describe quantitatively by aspect ratio) or
bulk structure (which we describe by the classifications of
'uniform' or 'boundary' disorder) in a sandpile induces the
presence of strong disorder in  avalanche statistics}.

Additionally we present, in Figure \ref{Fig5}, the mass-mass
correlation function of a particular disordered sandpile; the
curve has a peak, which indicates the average time between
avalanches. Since the avalanche size distribution for this
sandpile includes a preponderance of large events, we conclude
that the peak in the correlation function corresponds
approximately to the {\it time between large avalanches}. We  also
expect this timescale to manifest itself in the power spectrum of
the avalanches; and we expect it to vary strongly with the level
and nature of disorder in the sandpile. This work is in progress,
as are efforts to relate the timescale found above to a
characteristic spatial signature for large events.

We present in Figure \ref{Fig6} the normalised {\it equal-time}
height-height correlation function
$<dh(r+r_0)dh(r_0)>/<dh(r_0)^2>$ for a disordered sandpile. This
shows that the height deviations (from the instantaneous expected
column heights), in a disordered sandpile with $L=256$, are
positively correlated over about $80$ columns, but also have a
range where they are negatively correlated. In the inset we plot
the related function $1 - <dh(r+r_0)dh(r_0)>/<dh(r_0)^2> \equiv
<(dh(r+r_0)-dh(r_0))^2>/2<dh(r_0)^2>$. For separations $r$ much
less than the correlation length of the system, we should have
\cite{godreche}:
\begin{equation}
 <(dh(r+r_0)-dh(r_0))^2> \sim |r|^ {2 \alpha}
\end{equation}
and, therefore, we would expect the function in the inset of Figure
\ref{Fig6} to manifest  a similar $r$-dependence. A linear fit to
points with $r < 30$ (shown by the line in the inset of Figure
\ref{Fig6}) indicates a power-law dependence of the form $r^
{0.67}$ for $r << L$, implying that $\alpha \sim 0.34$. As we will
see below, this corresponds to the spatial roughening exponent of
an {\it ordered} sandpile. An explanation of this behaviour is
included in the next section.

\subsection {Spatial and temporal roughening of sandpile surfaces}
The hypothesis of dynamical scaling for sandpiles assumes that the
roughening process occurs in two stages. First, the surface
roughening is  time-dependent, Eq. (\ref{betaeqn}); then once the
roughness becomes temporally constant, the surface is said to
saturate, and all further deposition results in  surface
fluctuations governed by Eq. (\ref{alphaeqn}).

However, there is a subtlety concerning the first (i.e.
time-dependent) stage; "early"  model sandpiles are wedge-shaped
and the transition to saturation is accompanied by a gradual
build-up to a pile that has a single, sloping surface with a
suitable angle of repose.

We have taken this process into account to measure the dynamic
exponent $\beta$, Eq. (\ref{betaeqn}); in this case surface widths
are evaluated from the sloping portion of the pile. For the
roughening exponent $\alpha$, Eq. (\ref{alphaeqn}), we have
measured surface widths from mature piles that have only a sloping
surface.

Our results are:
\begin{itemize}
\item For disordered sandpiles ($L=2048$)
we find $\beta=0.42 \pm 0.05$; for ordered sandpiles ($L=2048$)
$\beta=0.17 \pm 0.05$.
\item For disordered sandpiles above a crossover
size of $L_c=90$ we find $\alpha=0.723 \pm 0.04$; while for
ordered piles we find $\alpha = 0.356 \pm 0.05$.
\item Based on the above values we find the dynamical
exponent $z$, has values of  $1.72 \pm 0.29$ and $2.09 \pm 0.84$
for the disordered and ordered sandpiles.
\end{itemize}

The variation of the surface width, $W$, as a function of $L$, is
shown in a log-log plot in Figure \ref{Fig7}. This figure shows
clearly the crossover in $\alpha$ as a function of system size,
for disordered sandpiles; the scaling behaviour of ordered
sandpiles is shown for comparison. Disordered sandpiles with sizes
below $L_c$ have $\alpha = 0.37 \pm 0.05$; this is in accord with
earlier work, \cite{amca} where the second peak in the avalanche
spectrum appeared only for disordered piles above crossover. The
existence of this crossover length has been variously interpreted
as a length related to reorganisation in the boundary layer of a
sandpile \cite{ambook} or to variations in the angle of repose in
a (disordered) sandpile \cite{jaegerprl}. Disorder appears to be
crucial for the existence of such experimentally observed
crossovers, since for example ordered models \cite{pacz},
\cite{hwa} show no crossover in their measurements of $\alpha$.
The crossover effect in a disordered sandpile is also indicated by
the height-height correlation function (Figure \ref{Fig6}). For
separations $r << L_c \sim 90$ in disordered sandpiles (with
length $L
>> L_c$), the exponent $\alpha$ obtained from the small $r$ behaviour
of the correlation function is that of the ordered sandpile. This
suggests that even in disordered sandpiles grains which are within
a crossover length, $L_c$, of each other tend to order i.e. the
examination of the height-height correlation function for
separations $r << L_c$ (Figure \ref{Fig6}) or the direct measurement of
$\alpha$ for system sizes $L << L_c$, as reported above, yields
the exponent of the ordered sandpile, $\alpha \sim 0.35$.

The above values indicate that while there is not a change of
universality class as one goes from an ordered to a disordered
sandpile ($z$ stays the same, within the error bars), the
disordered pile is clearly rougher with respect to both temporal
and spatial fluctuations ($\alpha$ and $\beta$ higher).

It is important to note that our measurements of surface exponents
are taken over many  realisations of the surfaces concerned. Thus,
even though, as demonstrated in earlier sections, the surface of a
disordered sandpile is temporarily smoothed by the propagation of
a large avalanche, it begins to roughen again as a result of
deposition; the values of $\alpha$ and $\beta$ that we measure are
averages over millions of such cycles and hence reflect the
roughening of the interface, in an average sense. By contrast, no
abnormally large events occur for the ordered sandpiles and this
is reflected by the lower values of fluctuations and exponents.

The most striking aspect of these exponents is that they indicate
that our present cellular-automaton model is a discrete version of
earlier continuum equations \cite{amcoupled1}, which were
formulated independently, to model the pouring of grains onto a
sloping surface. The exponents for our disordered pile are within
error bars, exactly those that were measured for the height
fluctuations of the surface in case 2, in ref. \cite{amcoupled1},
while those for the ordered pile are exactly those that were
measured for the fluctuations of the avalanches generated by the
mobile grains, in the same case.  This is in accord with the
notion that the avalanches which flow on an ordered pile generate
only mobile grains on the otherwise ordered surface, while as we
have demonstrated above, avalanches that flow on a disordered
pile, also {\it change the configuration of the surface} by
altering the distribution of height fluctuations (measured by the
surface widths). We are exploring these analogies further, but
note that this agreement is already a strong validation of both
models.
\section{Discussion and conclusions}
We have presented a thorough investigation of the effects of
avalanching on a sandpile surface, focusing on the
interrelationship between the nature of the avalanches and the
surfaces they leave behind. We have also postulated  a principle
of dynamical scaling for sandpile surfaces, and measured the
roughening exponents for a sample disordered sandpile. Finally, we
have related the characteristics of avalanching in our model
system to those obtained experimentally.

Our current investigations concern several questions left
unanswered  above. These include the dependence of the crossover
length $L_c$ on the disorder in the pile; as well as a fuller
investigation of the effect of the nature of disorder (i.e.
whether boundary or uniform). We would expect our correlation
functions to depend strongly on the nature and magnitude of the
disorder and we are undertaking a full quantitative study. Lastly,
we hope that an extension of the present analysis to  higher
dimensions will yield more extensive comparisons with experiments
than is presently available.

\acknowledgements GCB acknowledges support from the Biotechnology
and Biological Sciences Research Council, UK ($218$/FO$6522$).

\begin{table}
\caption{Instantaneous properties of disordered model sandpiles}
\label{Table1}
\begin{tabular}{|c|c|c|c|c|}
$L$&State of pile&Packing fraction $\phi$&Slope&Width\\ \hline
\hline 256&Before&0.997&1.15&4.45\\ \hline
256&After&1.000&1.07&2.06\\ \hline 64&Before&0.991&1.17&1.41\\
\hline 64&After&0.998&1.04&1.15\\
\end{tabular}
\end{table}

\begin{table}
\caption{Properties of model sandpiles}
\label{Table2}
\begin{tabular}{|c|c|c|c|}
Aspect ratio&Packing fraction $\phi$&Slope&Width\\ \hline \hline
0.6&0.997&1.44&2.33\\ \hline 0.65&0.997&1.42&2.34\\ \hline
0.7&0.997&1.12&3.76\\ \hline 0.75&0.997&1.19&3.74\\ \hline
0.8&0.998&1.17&2.35\\ \hline 0.9&0.999&1.25&2.29\\ \hline
0.95&1.000&1.32&2.37\\ \hline 1.0&1.000&1.40&2.41\\
\end{tabular}
\end{table}

\begin{figure}
\caption{(a) A time series of the mass for a model sandpile
($L=256$) that has been built to include a surface layer
containing structural disorder (Mass is measured in lattice grain
units). The vertical line indicates the position in this series of
the large avalanche illustrated in Figure \ref{Fig1} c, d. (b) A
log-log plot of the event size distribution for a model sandpile
($L=256$) that includes a surface layer containing structural
disorder. (c) An illustration of a large wedge shaped avalanche in
a model sandpile ($L=256$) that has been built to include a
surface layer containing structural disorder. A lighter aftermath
pile has been superposed onto the dark precursor pile and an arrow
shows the point at which the event was initiated. The inset shows
the relative positions of the two surfaces and their relationship
to a pile that has a smooth slope (Height and position are
measured in lattice units). (d) A detailed picture of the internal
structure of a model sandpile in the aftermath of a large
avalanche event. The individual grains of the aftermath pile (for
columns $1-128$ of a sandpile with $L=256$) are superposed on the
gray outline of the precursor pile.} \label{Fig1}
\end{figure}

\begin{figure}
\caption{(a) A time series of the mass for a model sandpile
($L=256$) that has been built to exclude any structural disorder
(Mass is measured in lattice grain units). The vertical line
indicates the position in the series of the avalanche illustrated
in Figure \ref{Fig2} c, d. (b) A log-log plot of the event size
distribution for a model sandpile ($L=256$) that excludes
structural disorder. (c) An illustration of a large 'flat'
avalanche in a model sandpile ($L=256$) that excludes structural
disorder. A lighter aftermath pile has been superposed onto the
dark precursor pile and an arrow shows the point at which the
event was initiated. The inset shows the relative positions of the
two surfaces and their relationship to a pile that has a single
smooth slope (Height and position are measured in lattice units).
(d) A detailed picture of the internal structure of a model
sandpile in the aftermath of a large avalanche event. The
individual grains for the aftermath pile (for columns $1-128$ of
an ordered pile with $L=256$) are superposed on the grey outline
of the precursor pile.} \label{Fig2}
\end{figure}

\begin{figure}
\caption{ (a) A time series of the mass for a model sandpile
($L=64$) that has been built to include a surface layer containing
structural disorder (Mass is measured in lattice grain units). The
vertical line indicates the position in this series of the large
avalanche illustrated in Figure \ref{Fig3} c, d. (b) A log-log
plot of the event size distribution for a model sandpile ($L=64$)
that includes a surface layer containing structural disorder. (c)
An illustration of a large wedge shaped avalanche in a model
sandpile ($L=64$) that has been built to include a surface layer
containing structural disorder. A lighter aftermath pile has been
superposed onto the dark precursor pile and an arrow shows the
point at which the event was initiated. The inset shows the
relative positions of the two surfaces and their relationship to a
pile that has a smooth slope (Height and position are measured in
lattice units). (d) A detailed picture of the internal structure
of a model sandpile in the aftermath of a large avalanche event.
The individual grains of the aftermath pile are superposed on the
gray outline of the precursor pile.} \label{Fig3}
\end{figure}

\begin{figure}
\caption{Exit mass event size distributions for a disordered
sandpile model with $L=256$, and $a=  \text{a) } 0.6, \text{b) }
0.7, \text{c) } 0.75, \text{d) } 0.85, \text{e) } 1.0$. The curves
have each been shifted by $0.5$ to make them distinct.}
\label{Fig4}
\end{figure}

\begin{figure}
\caption{The mass-mass correlation function for a disordered model
sandpile with $L=256$.} \label{Fig5}
\end{figure}

\begin{figure}
\caption{The normalized correlation function of column height
deviations for a disordered model sandpile with $L=256$. The inset
illustrates an initial decay with $r$, going as $r^{-0.67}$ ($r$
is measured in lattice units).} \label{Fig6}
\end{figure}

\begin{figure}
\caption{A log-log plot of the surface widths, $W$, against the
system size, $L$, for model sandpiles; (x) ordered piles, (o)
disordered piles. Widths are measured in lattice units and all
points have error bars that are $\sim0.02$.} \label{Fig7}
\end{figure}


\begin{references}
\bibitem{held}
H. M. Jaeger, C. Liu and S. R. Nagel, Phys. Rev.
Lett. {\bf 62},
40 (1989);
G. A. Held, D. H. Solina, D. T. Keane, W. J. Haag,
P. M. Horn, and G. Grinstein, Phys. Rev. Lett. {\bf 65}, 1120
(1990).
\bibitem{jaegerRMP} H. M. Jaeger, S. R. Nagel and R. P. Behringer, Rev.
Mod. Phys. {\bf 68}, 1259 (1996).
\bibitem{daerr} Adrien Daerr and St\'ephane  Douady, Nature {\bf 399},
241 (1999).
\bibitem{ball} R. E. Snyder and R. C. Ball, Phys. Rev. E {\bf 49}, 104
(1994).
\bibitem{amcoupled1} Anita Mehta, J. M. Luck and R. J. Needs, Phys. Rev.
E {\bf 53}, 92 (1996).
\bibitem{amcoupled2} P. Biswas, A. Majumdar, Anita Mehta, and J. K. Bhattacharjee,
Phys. Rev. E {\bf 58}, 1266 (1998).
\bibitem{amca} Anita Mehta and G. C. Barker, Europhysics Lett. {\bf 27},
501 (1994); Anita Mehta, G.
C. Barker, J. M. Luck and R. J. Needs, Physica A {\bf 224}, 48
(1996).
\bibitem{kadanoff} L. P. Kadanoff, S. R. Nagel, L. Wu and S.-M Zhou,
Phys. Rev. A {\bf 39}, 6254 (1989); P. Bak, C. Tang and K.
Wiesenfeld, Phys. Rev. A {\bf 38}, 368 (1988).
\bibitem{ambook} Anita Mehta, Physica A {\bf 186} 121, (1992); Anita
Mehta, in {\it Granular Matter: An Interdisciplinary Approach }, 1, ed. Anita Mehta,
(Springer-Verlag, New York, 1994).
\bibitem{caparam} Throughout this paper we refer to disordered sandpiles
described in reference \cite{amca} with parameters $z_0=2, z_1=20$ and $a =
0.7$, unless otherwise stated.
\bibitem{ampune} Anita Mehta in {\it Structure and dynamics of materials in
the mesoscopic domain},
eds. M. Lal et al., 340, (Imperial College Press and the Royal Society, London, 1999).
\bibitem{krug} J. Krug, Adv. Phys. {\bf 46}, 1 (1997).
\bibitem{amunpub} G. C. Barker and Anita Mehta (unpublished).
\bibitem{ampre} G. C. Barker and Anita Mehta Phys. Rev. E {\bf 53}, 5704
(1996).
\bibitem{amprl} Anita Mehta and G. C. Barker, Phys. Rev. Lett. {\bf 67},
394 (1991).
\bibitem{nagel} J. B. Knight, C. G. Fandrich, C. N. Lau, H. M. Jaeger
and S. R. Nagel, Phys. Rev. E {\bf 51}, 3957 (1995); E. R. Nowak, J. B.
Knight, E. Ben-Naim, H. M. Jaeger and S. R. Nagel, Phys. Rev. E
{\bf 57}, 1971 (1998).
\bibitem{frette} V. Frette, K. Christensen, A. Malthe-Sorenssen, J.
Feder, T.
Jossang and P. Meakin, Nature {\bf 379}, 49 (1996).\bibitem{godreche} L. Sander,
in {\it Solids far from Equilibrium },
ed. C.
Godr\`eche, (Cambridge University Press, Cambridge, 1991).
\bibitem{jaegerprl}
S. R. Nagel, Rev. Mod. Phys. {\bf 64}, 321 (1992).
\bibitem{pacz} M. Paczuski and S. Boettcher, Phys. Rev. Lett. {\bf
77}, 111, (1996).
\bibitem{hwa} T. Hwa and M. Kardar, Phys. Rev. A {\bf 45}, 7002,
(1992).
\end{references}
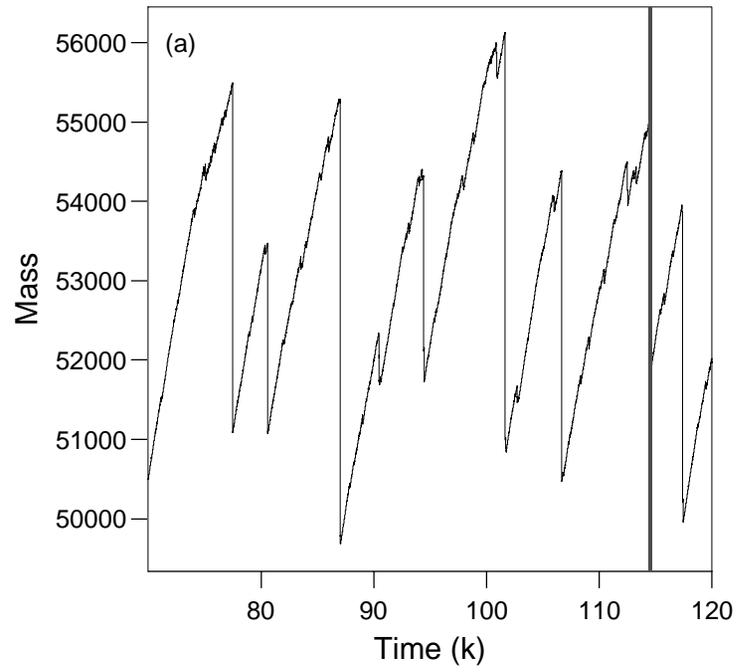
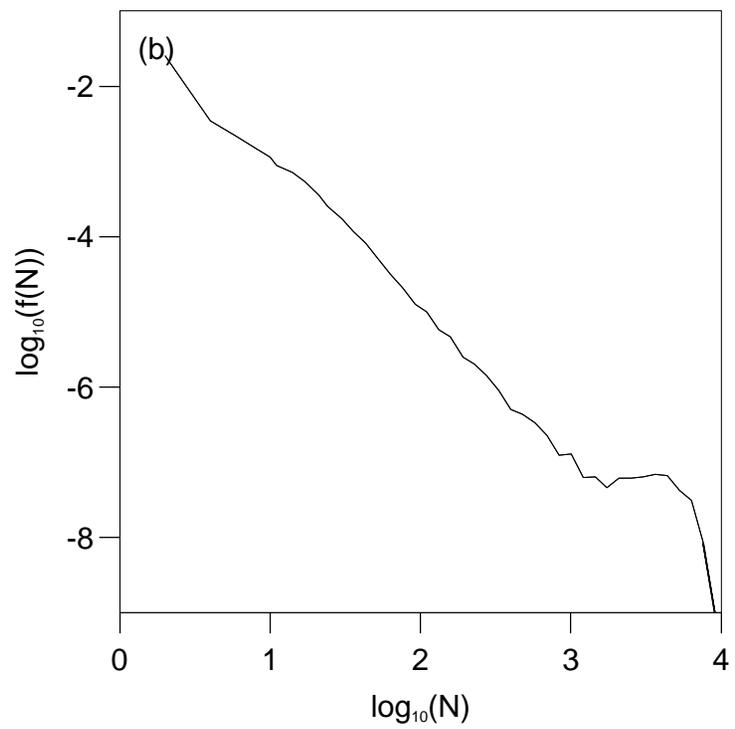
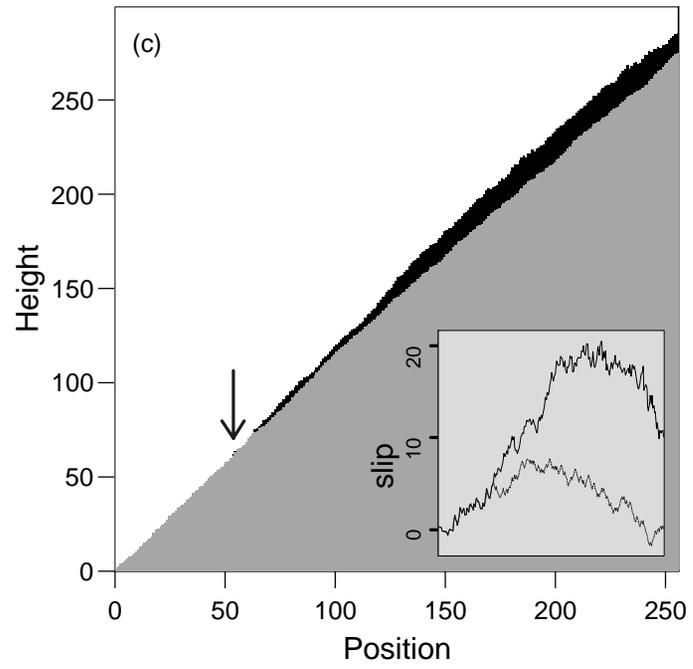
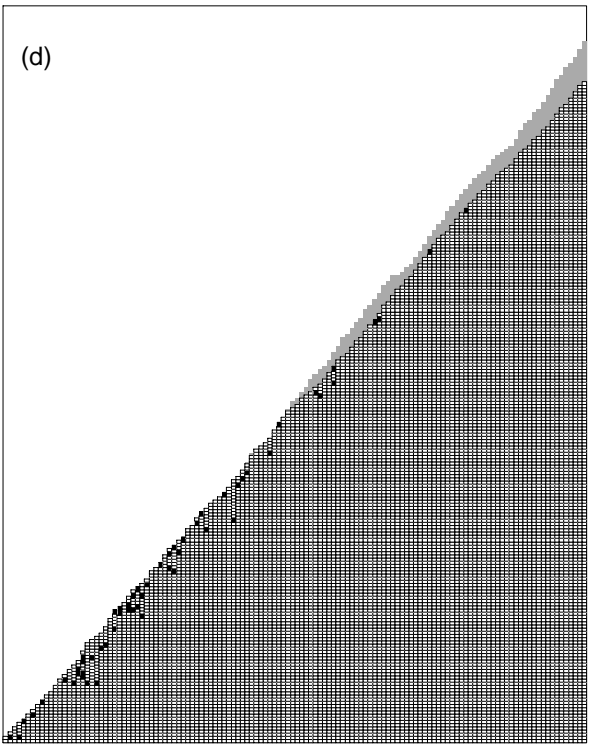
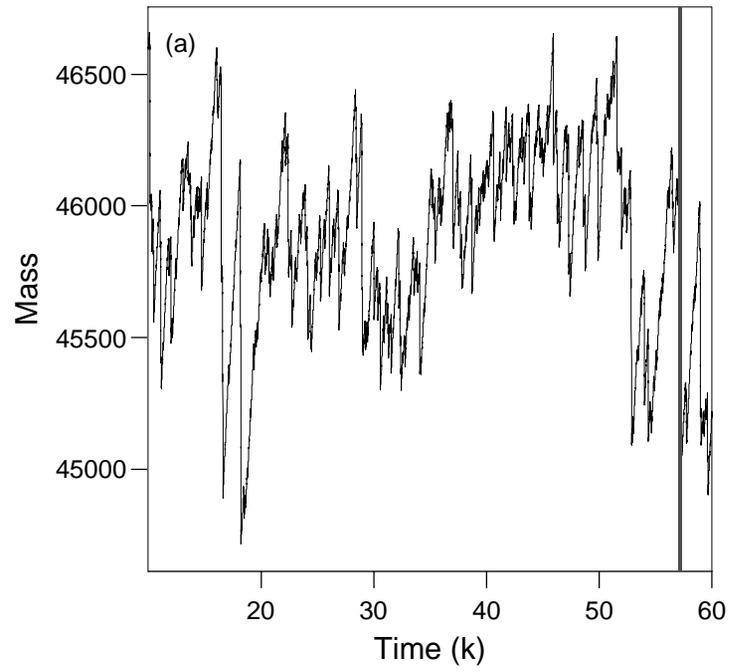
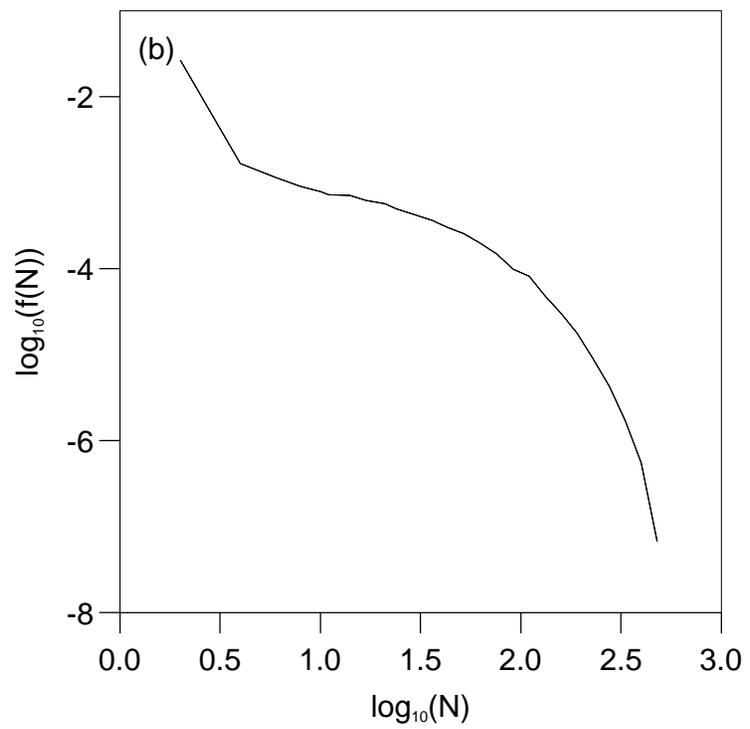
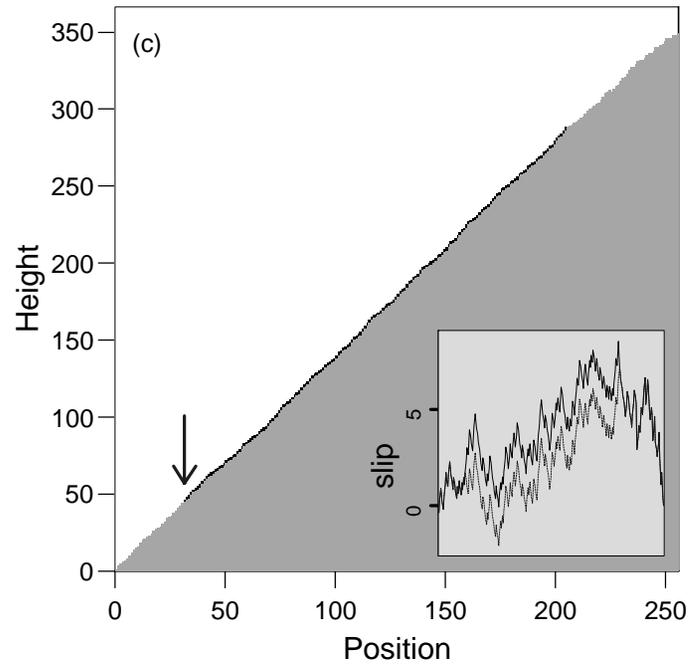
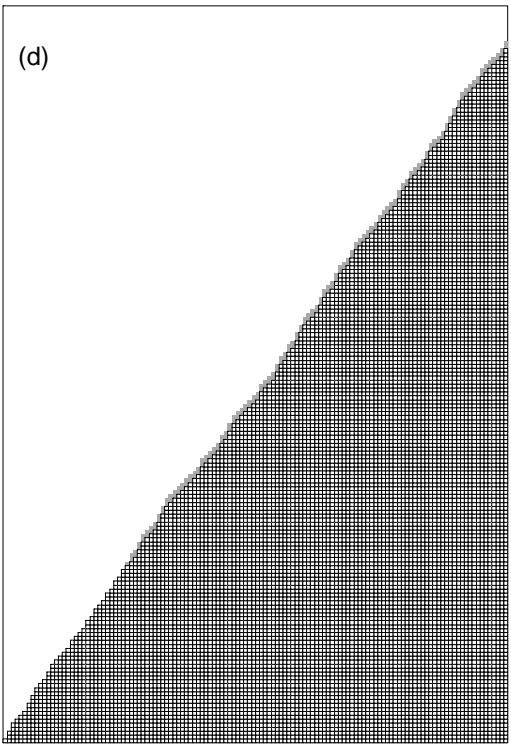
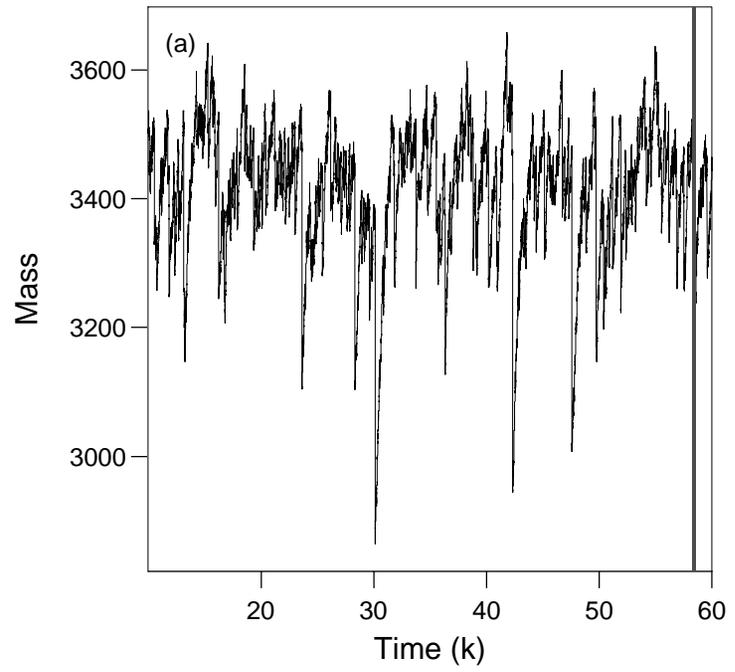
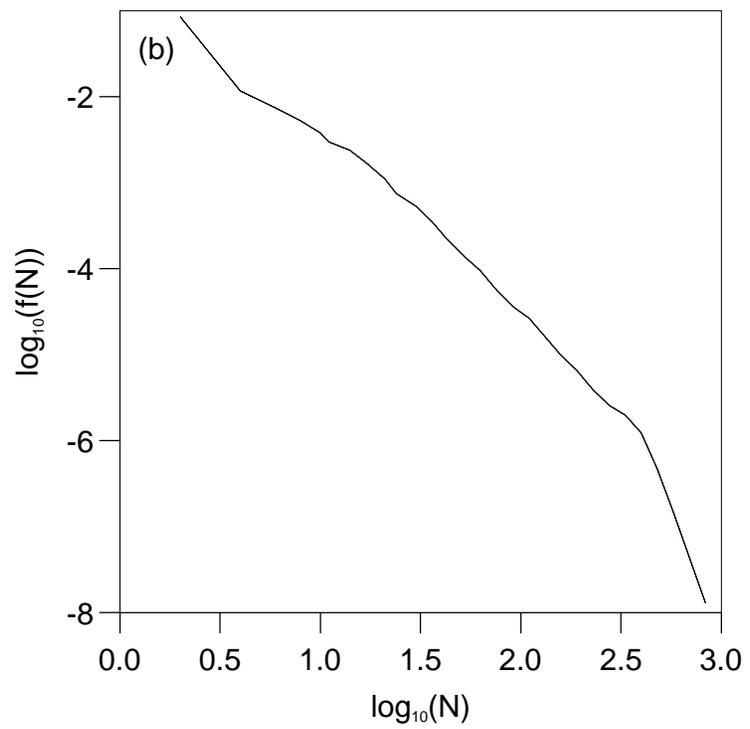
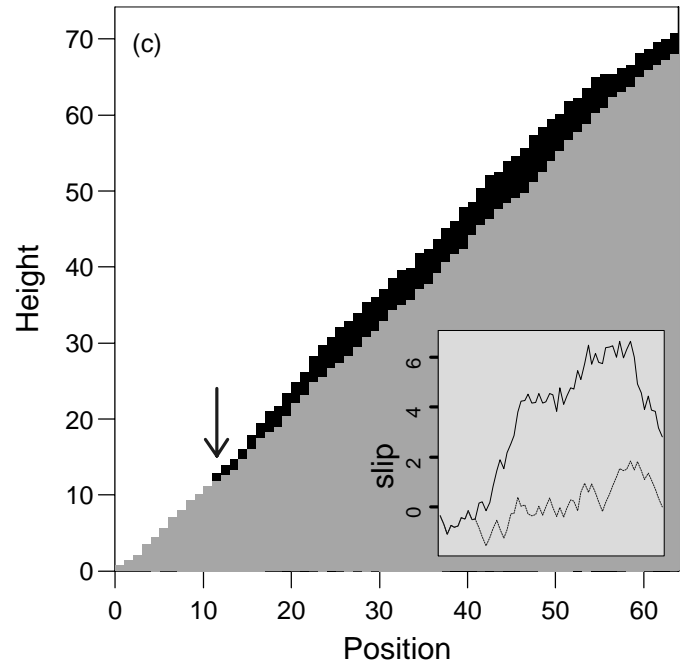
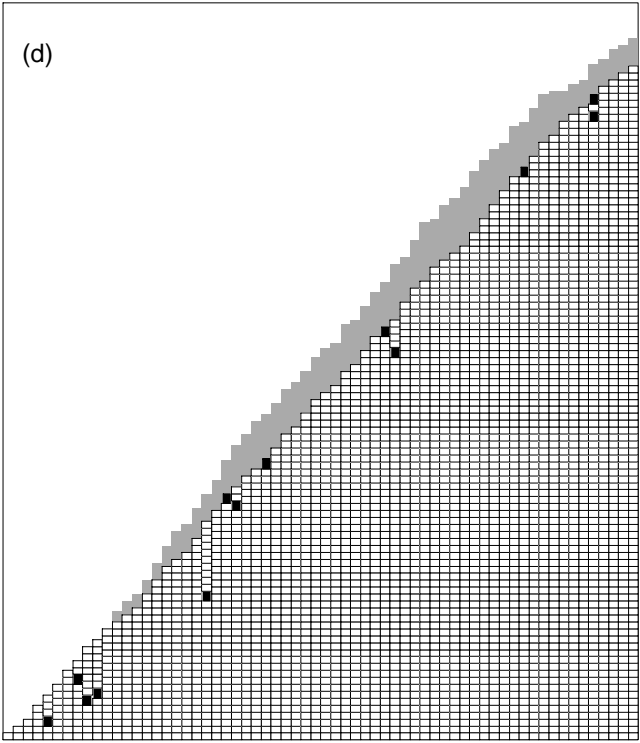
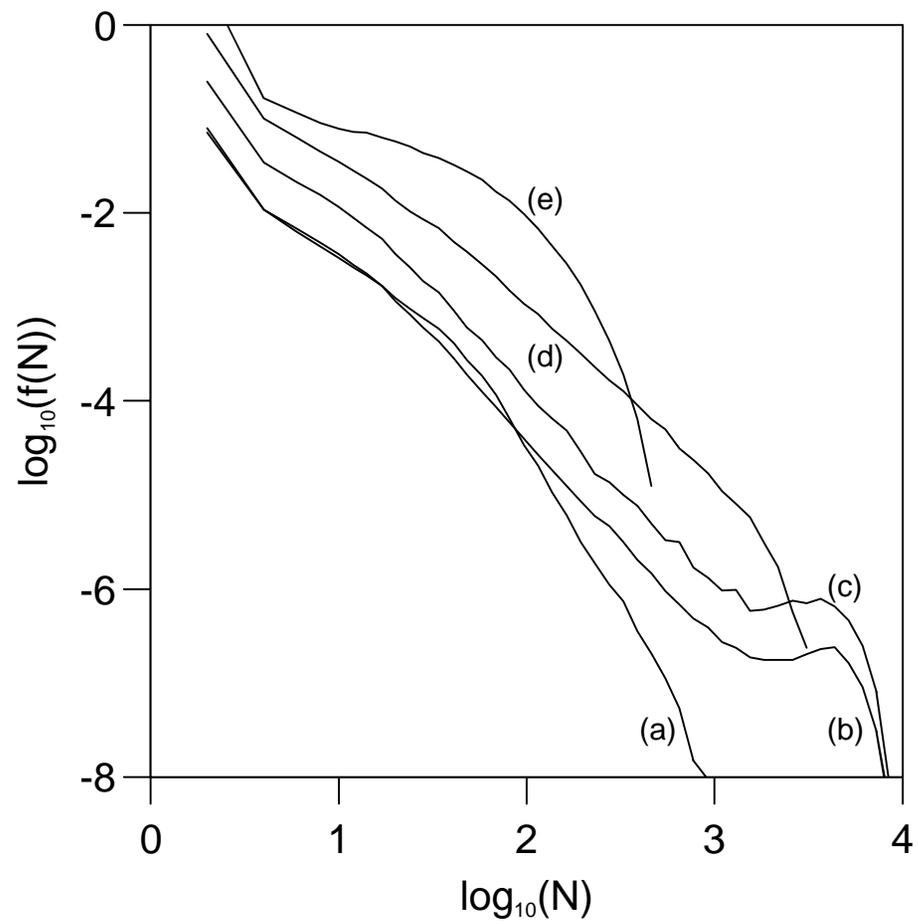
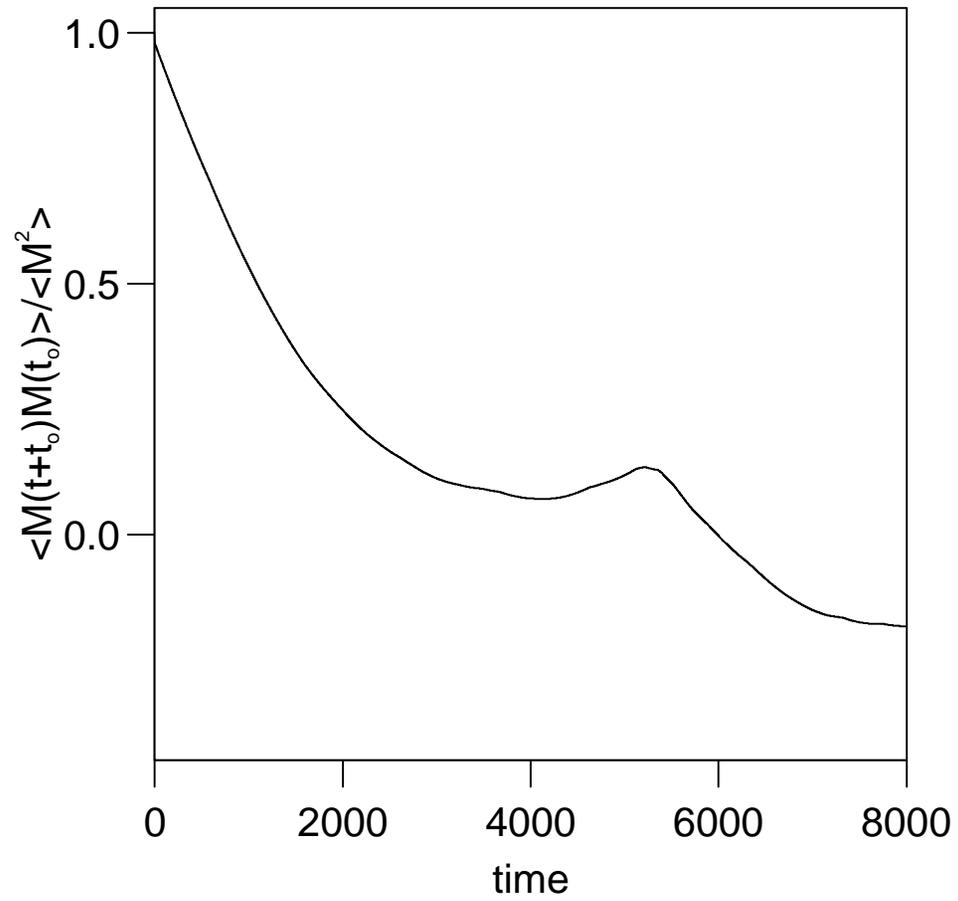
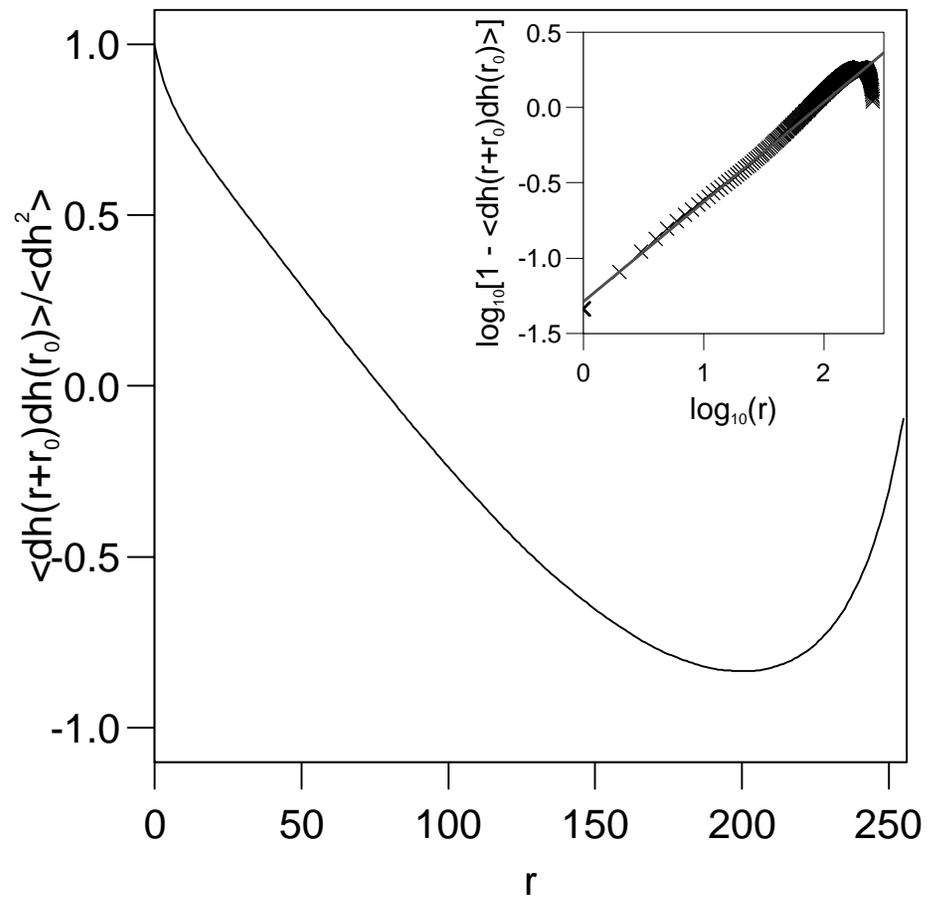
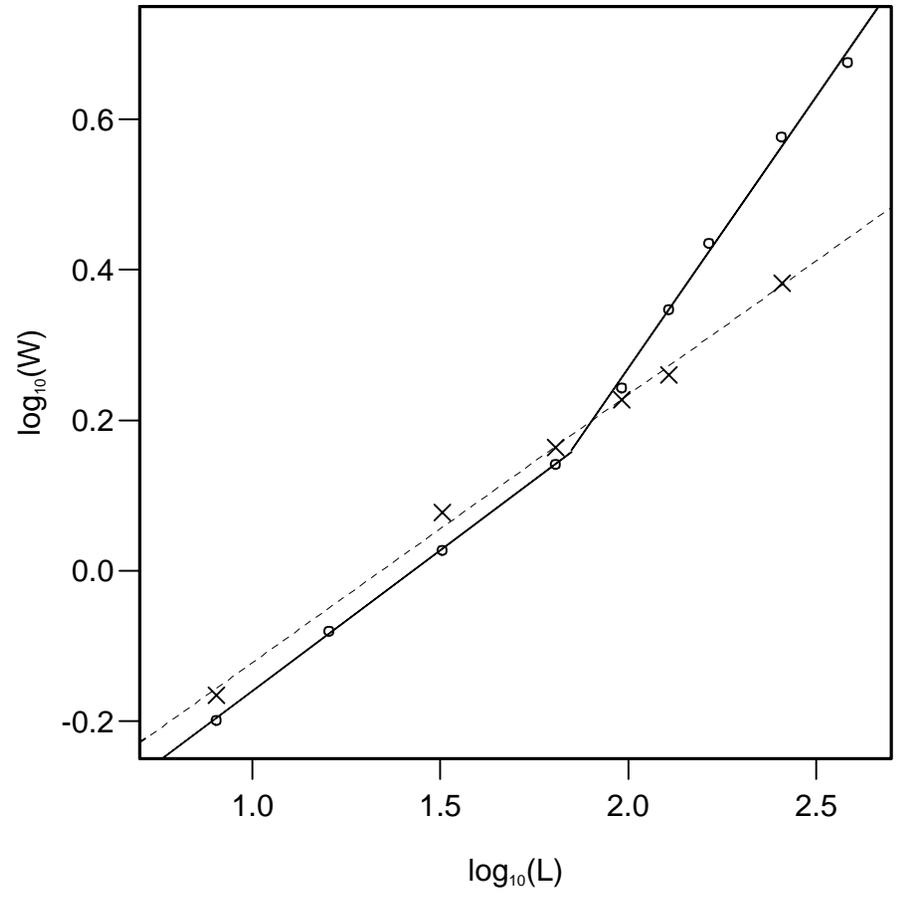
\end{document}